\documentclass[a4paper]{panl}
\usepackage{cite}
\usepackage{wrapfig}
\usepackage{graphicx}
\usepackage{amssymb}
\usepackage{amsfonts}
\usepackage{amsmath}
\usepackage{longtable}
\usepackage{rotating}
\usepackage{lscape}
\usepackage{epsfig}
\usepackage{multirow}

\usepackage{hepnames}

\newcommand{\el}{\mbox{e$^{-}$}}

\newcommand{\pos}{\mbox{e$^{+}$}}
\newcommand{\pbar}{\mbox{$\overline{\mathrm p}$}}
\newcommand{\Hbar}{\mbox{$\overline{\mathrm H}$}}

\newcommand{\mprot}{$m_{\mathrm{p}}$}

\newcommand{\nus}{$\nu_{\sigma}$}
\newcommand{\nupone}{$\nu_{\pi_1}$}

\newcommand{\nuS}{$\nu_{\mathrm{1S-2S}}$}
\newcommand{\nuSH}{$\nu_{\mathrm{1S-2S}}^{\mathrm{H}}$}
\newcommand{\nuSHbar}{$\nu_{\mathrm{1S-2S}}^{\overline{\mathrm{H}}}$}
\newcommand{\nuLS}{$\nu_{\mathrm{2S-2P}}$}

\newcommand{\nuLSHbar}{$\nu_{\mathrm{2S-2P}}^{\overline{\mathrm{H}}}$}
\newcommand{\nuHFS}{$\nu_{\mathrm{HFS}}$}

\newcommand{\nuHFSHbar}{$\nu_{\mathrm{HFS}}^{\overline{\mathrm{H}}}$}

\newcommand{\Bext}{\mbox{$B_{\mathrm{ext}}$}}

\originalTeX
\begin{document}

\title{Hyperfine spectroscopy of antihydrogen, hydrogen, and deuterium}
\maketitle
\authors{E.\,Widmann$^{a,}$\footnote{E-mail: eberhard.widmann@oeaw.ac.at} for the ASACUSA Cusp collaboration
}

\setcounter{footnote}{0}
\from{$^{a}$\,Stefan Meyer Institute, Austrian Academy of Sciences, Kegelgasse 27, 1030 Vienna, Austria}

\begin{abstract}
The prospects of tests of CPT symmetry using precision spectroscopy of antihydrogen are discussed with special emphasis on the ground-state hyperfine structure, a measurement of which is the aim of the ASACUSA collaboration at the AD/ELENA facility of CERN. Ongoing parallel experiments using hyperfine spectroscopy of hydrogen and deuterium aiming at studying Lorentz invariance by determining coefficients of the Standard Model Extension framework are described.
\end{abstract}
\vspace*{6pt}


\label{sec:intro}
\section*{Introduction}

Antihydrogen is one of the most sensitive tools to study CPT symmetry, since it is the simplest anti-atom and its CPT conjugate counterpart, hydrogen, is one of the most precisely studied atomic systems. Antihydrogen is produced at the Antiproton Decelerator (AD) of CERN since 2000, and its extension to lower energies ELENA, which has just started operation in fall of 2021. The ASACUSA collaboration  proposed to measure the ground-state hyperfine splitting of antihydrogen at the first PSAS conference in 2000 \cite{WidmannEtAl2001}. The first production of antihydrogen in ASACUSA was achieved in 2010 \cite{Enomoto:2010uqcoll}, and the first beam of antihydrogen (\Hbar) in a field-free region was observed in 2014 \cite{Kuroda:2014fkcoll}. Since then efforts are ongoing to increase the \Hbar\ production rate and the ground state population. A first measurement showed, as expected from the three-body recombination mechanism used, a population of states of predominantly high principle quantum numbers \cite{Kolbinger:2021coll}. In parallel, ASACUSA is operating a cold hydrogen beam for hyperfine spectroscopy, where the ground-state hyperfine structure of hydrogen was measured with 2.7 ppb precision\footnote{In the following,
precision stands for, unless otherwise stated, relative precision.} \cite{Diermaier:2017}.

\section*{Tests of CPT symmetry}

\begin{figure}[]
	\begin{center}
	    \includegraphics[width=100mm]{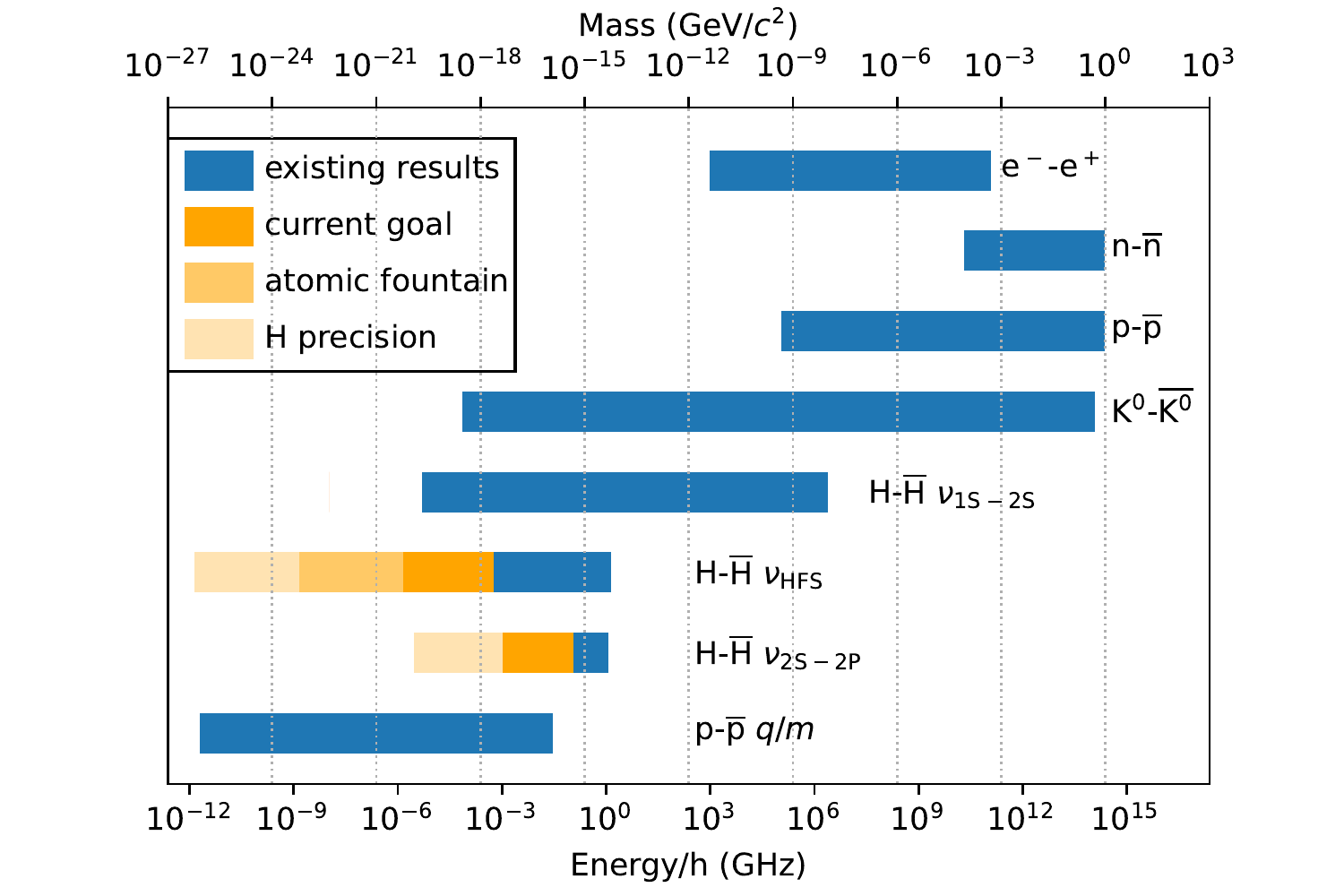}
		\vspace{-3mm}
		\caption{\label{fig:CPT} Comparison of several tests of CPT symmetry on an absolute energy scale. Bar's right edge: value of measured quantity, 
		left edge: absolute sensitivity. The bar length corresponds to the relative precision of the measurements. 
		}
	\end{center}
	\vspace{-5mm}
\end{figure}

Studies of CPT comparing particle properties have been performed for a long time, an overview of measurements is published by the Particle Data Group PDG \cite{Zyla:2020zbs} and a selection is shown in Fig.~\ref{fig:CPT}. Here we focus on measurements of masses and frequencies, since these quantities can easily be converted into each other. 
In recent years, first results of laser spectroscopy of antihydrogen have been obtained by the ALPHA collaboration at the AD. Fig.~\ref{fig:CPT} includes the measured quantities for the three most relevant transitions: the 1S--2S two-photon transition \nuS, the Lamb shift \nuLS, and the ground-state hyperfine splitting \nuHFS. \nuS\ and \nuHFS\ are the most precisely measured transitions in hydrogen (\nuS\ with relative precision of $\delta \nu = \Delta \nu/\nu = 4.2\times 10^{-15}$ \cite{Parthey:2011ys,Matveev:2013},  \nuHFS\ with  $\delta \nu = 7\times10^{-13}$ in a hydrogen maser \cite{Karshenboim:2005wz}). \nuLS\ has recently been measured with improved precision of $\delta \nu = 3\times10^{-6}$ by the group of E. Hessels \cite{Bezginov:2019}. Here the precision is limited by the short lifetime of the 2P state leading to a width of $\Gamma_ \mathrm{2P}\sim 100$ MHz. 

Experimental results of \Hbar\ are shown 
in Fig.~\ref{fig:CPT}: \nuSHbar\ was measured to $\delta \nu = 2\times10^{-12}$ \cite{Ahmadi:2018acoll}, although  the measurement is taken in a background magnetic field of $B=1.033$ T and compared to the calculated value of \nuSH $(B=1.033$ T) by using QED. As the quantity measured is dominated by the electro-magnetic interaction, this makes the current result not a pure test of CPT which would require comparing \nuSH\ and \nuSHbar\ at the same magnetic field.  \nuLSHbar\ was determined to $\delta \nu = 11$\% \cite{Ahmadi:2020coll} from the measured frequencies of \nuSHbar\ and 
$\nu_{\mathrm{1S-2P}}^{\overline{\mathrm{H}}}$.  
The zero-field value of \nuHFSHbar\ was directly determined to $\delta \nu = 4\times 10^{-4}$ \cite{Ahmadi:2017bcoll}  using the difference of two microwave transitions. In addition, the estimated precision of planned experiments is indicated in the figure. \nuHFS: current goal: first goal for in-beam measurement of ASACUSA, atomic fountain: line width of an atomic fountain experiment, H precision: hydrogen maser result. \nuLS: current goal: estimated achievable accuracy \cite{Crivelli:16}, H precision: accuracy for hydrogen. \pbar\ charge-to-mass ratio \cite{Ulmer:2015}, note that in this case only the left edge is well defined, while the right edge is less precise since the cyclotron frequency is proportional to the magnetic field in which the measurement is taken.

While the 1S--2S transition is caused by the electromagnetic interaction, the hyperfine structure is caused by the magnetic interaction of the \el/\pos\ and p/\pbar\ magnetic moments. 
The leading order term deviates from experiment due to  structure dependent corrections (cf. Table~\ref{tab:Zemach}), which have a magnitude of $\sim 33$ ppm. For their determination, experimental values for the electric and magnetic form factors of the proton obtained by electron scattering are needed. Ref. \cite{Carlson:2008} arrives at a remaining difference between theory and experiment of $0.86(78)$ ppm. In summary, since the equality of magnetic moments of \el/\pos\ and p/\pbar\ are known to $10^{-10}$ \cite{Zyla:2020zbs} and $10^{-9}$ \cite{Smorra:2017} precision, resp.,  a measurement of \nuHFSHbar\ with precision of 1 ppm (which is the first goal of ASACUSA) will determine the magnetic sub-structure of the antiproton to a few \%. A similar accuracy for the electric charge radius of the antiproton is predicted from a planned Lamb shift measurement for \Hbar\ \cite{Crivelli:16}. 

\begin{table}[t]
    \centering
    \caption{Finite size effects in the hyperfine structure of hydrogen \cite{Carlson:2008}.}
    \medskip
    \addtolength\tabcolsep{-0.8mm}
    \begin{tabular}{l|r}
    \hline
    Deviation from $\nu_F$ due to proton structure    &  $-32.77(1)$ ppm\\ \hline
    Recoil corrections   &  +5.85(7) ppm \\
    Finite electric and magnetic radius (Zemach corrections)     & $-41.43(44)$ ppm\\
    Polarizability of proton  &  $+1.88(64)$ ppm\\ \hline
    Remaining deviation theory-experiment & $+0.86(78)$ ppm\\ \hline
    \end{tabular}
    \label{tab:Zemach}
\end{table}

\section*{Hyperfine structure and the Standard Model Extension}

The group of V.A. Kostelecky at Indiana University has
developed an extension to the standard model (SME -- Standard Model Extension) that includes both
CPT as well as Lorentz-invariance violating (LIV) terms in the
Lagrangian of a quantum field theory. First a minimal SME was developed which uses operators up to mass dimension 4
\cite{Kostelecky:1995e,Colladay:1997vn,Colladay:1998tw}.
The modified Dirac-equation of the minimal SME has the following structure:
\begin{equation}
(i \gamma^{\mu}D_{\mu} - m_e - a_{\mu}^e \gamma^{\mu}
- b_{\mu}^e \gamma_{5} \gamma^{\mu} -
\frac{1}{2} H^e_{\mu\nu}\sigma^{\mu\nu} + i c^e_{\mu\nu} \gamma^{\mu}D^{\nu}
+i d^e_{\mu\nu} \gamma_{5}\gamma^{\mu}D^{\nu})\Psi =0,
\end{equation}
with $i D_{\mu} \equiv i \partial_{\mu}-q A_\mu $, $q=-|e|$ is the charge and $A_\mu$ the electro-magnetic potential. The additional terms containing $a_{\mu}^e$ and  $b_{\mu}^e$ violate both
Lorentz-invariance  and CPT, while the ones containing $H^e_{\mu\nu}$,
$c^e_{\mu\nu}$ and $d^e_{\mu\nu}$ violate only Lorentz-invariance. The upper index $e$ stands  for the electron, implying that a set of parameters exists for each particle under investigation.

\begin{figure}[]
	\begin{center}
		\includegraphics[width=100mm]{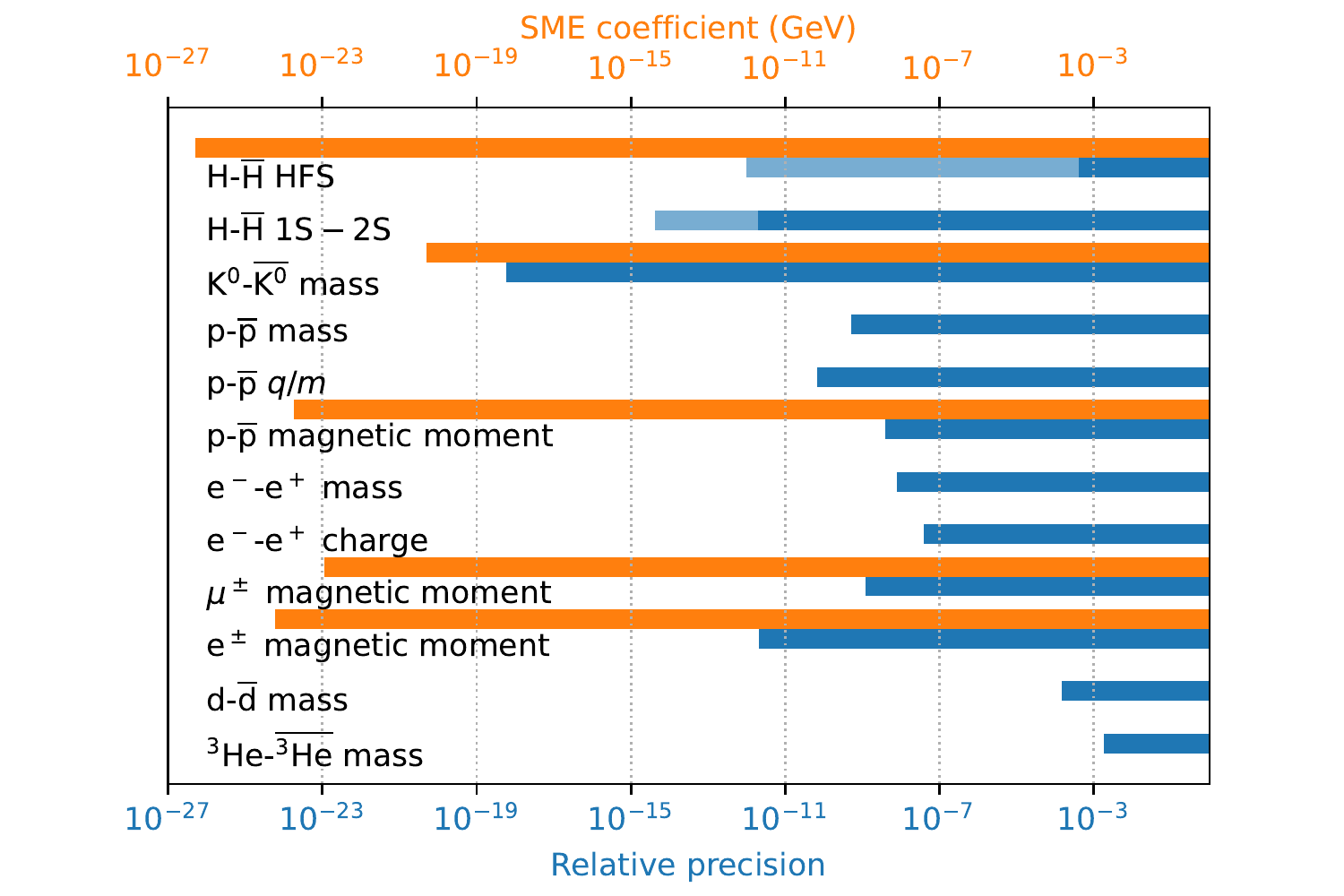}
		\vspace{-3mm}
		\caption{\label{fig:SME} Relative experimental precision and corresponding SME coefficients for various quantities, Dark blue: existing experimental values for both H and \Hbar, light blue: experimental precision for H, orange: SME coefficient. SME values do not exist for mass comparisons, \nuS\ is not sensitive to CPT in the minimal SME.}
	\end{center}
	\vspace{-5mm}
\end{figure}

It is important to note that the so-introduced symmetry-violating coefficients have the dimension of mass or energy.  This  results in the conjecture that if comparing CPT tests of different particles or different particle properties, what counts is the {\em absolute} precision in eV and not the  {\em relative} precision.

Although this model does not directly predict any CPT violation
nor LIV, it can be used as basis to compare CPT tests in different
sectors, and as a guide where to look for possible CPT violating
effects. In fact, various groups have already done so, their results are summarized in the regularly updated ``Data Tables for Lorentz and CPT Violation'' of \cite{Kostelecky:2011jr,Kostelecky:2021jr}.

Fig.~\ref{fig:SME}   compares various tests of CPT. The experimental values  are from PDG \cite{Zyla:2020zbs} except for \mprot\ \cite{Hori:2016} and the values for the antideuteron and anti-$^3$He masses \cite{Adam:2015coll}. The SME coefficients are taken from the latest version of the data tables \cite{Kostelecky:2011jr}. In the case of \nuHFSHbar, the measurement by ALPHA \cite{Ahmadi:2017bcoll} uses a combination of transitions which is not sensitive in the SME, but results from siderial variations of a hydrogen maser \cite{Phillips:01,Humphrey:2003} exist which set limits on a combination of SME coefficients $\tilde{b}_3^w = b_3^w - d_{30}^w  m_w - H_{12}^w$ with $w=e,p$. For antihydrogen, the coefficients $d$ and $H$ reverse sign, while $b$ do not. A measurement of \nuHFSHbar\ will therefore be able to separate the different contributions to $\tilde{b}$.

\section*{Hyperfine spectroscopy of hydrogen and deuterium}

Initially meant as a verification of the hyperfine spectroscopy method of ASACUSA, an atomic hydrogen beam was constructed having the expected \Hbar\ velocity \cite{Malbrunot:2019coll} and, using the same microwave cavity and supercondcuting sextupole built for the antihydrogen experiment,  the zero-field hyperfine splitting of H was determined with 4 Hz (2.7 ppb) precision \cite{Diermaier:2017}. The quantity measured was the so-called $\sigma$-transition ($(F, m_F)=(1,0)\rightarrow(0,0)$, $F$: total spin, $m_F$  magnetic quantum number), which in the SME is not sensitive to CPT \cite{Bluhm:1999vq}. In order to investigate the CPT-sensitive $\pi_1$-transition ($(1,1)\rightarrow(0,0)$), a modified setup was built  \cite{Malbrunot:2017coll} and first results were obtained with 10--30 ppb precision \cite{Widmann:2019coll}. 

In 2013, the minimal SME was extended to include operators of arbitrary dimensions \cite{KosteleckyEtAl2013}. This leads to new possibilities in the spectroscopy of hydrogen, where coefficients exist that depend on the orientation of the externally applied static magnet field \Bext\ \cite{Kostelecky:2015}. As described in \cite{Malbrunot:2017coll}, ASACUSA started measurements inverting the direction of \Bext. The experiment is ongoing, first results are expected in early 2022. The largest systematic error is the reproducibility of the absolute value of \Bext\ when inverting its direction, which can be controlled by measuring both transition frequencies \nus\ and \nupone\ since \nus\ is not shifted in the SME. Due to the limited sensitivity of \nus(\Bext) at \Bext $\sim$ Gauss, the current error on $\delta \nu_\pi$=\nupone($\boldsymbol{B}_\mathrm{ext}) - $\nupone($-\boldsymbol{B}_\mathrm{ext}$) is about 300 Hz (1$\sigma$). Higher sensitivities can be reached by either operating at much larger \Bext\ or using other methods to measure its value than \nus.

Within the non-minimal SME, there are also new effects predicted in the hyperfine structure of deuterium. Most interesting are those depending on powers of the relative momenta of proton and neutron  in the deuterium nucleus, where the proton has a Fermi momentum of $\sim 100$ MeV/$c$. The sensitivity of coefficients  is enhanced $10^9$ times or $10^{18}$ times, resp., depending on the power of the momentum. ASACUSA is preparing a measurement with a deuterium beam and a cavity for $\nu_\mathrm{HF} =300\ldots400$ MHz to measure several hyperfine transition frequencies in the coming year.

\section*{Summary}

As shown in the previous sections, the hyperfine spectroscopy of antihydrogen, hydrogen, and deuterium can provide highly sensitive tests of CPT symmetry and Lorentz invariance. First results from ASACUSA are expected within the next few years for \nuHFSHbar, and earlier for hydrogen and deuterium.


\section*{Acknowledgements}
The authors thanks his colleagues of the ASACUSA collaboration for the continuous cooperation and many fruitful discussions, in particular T.~Yamazaki, as well as extensive discussions with S. Karshenboim, V.A. Koste\-lecky, and R. Lehnert. The work  was supported 
by the Austrian Science Fund (FWF): W1252-N27, the Grant-in-Aid for Specially Promoted Research 24000008 of Japanese MEXT, Special Research Projects for Basic Science of RIKEN, and Universit\`a di Brescia and Istituto Nazionale di Fisica Nucleare. 



\bibliographystyle{pepan}
\bibliography{hbar-physrep,ffk2021}

\end{document}